\documentclass[12pt]{article}
\usepackage[margin=-1.0in]{geometry}
\usepackage{latexsym,amsmath}
\usepackage[dvips]{graphicx}
\usepackage{lscape}
\usepackage{color}
\usepackage[pagewise]{lineno}

\usepackage{graphicx} 
\usepackage {epstopdf}
\usepackage {subfigure}
\usepackage{hyperref}
\usepackage{booktabs} 
\usepackage{amsmath}
\usepackage{multicol}
\usepackage{multirow}
\usepackage{xcolor}
\usepackage{float}
\usepackage{amsthm}
\usepackage{amsfonts}

\newtheorem{theorem}{Theorem}

\def\doublespace{\baselineskip=24pt}

\newcommand{\uiota}             {\mbox{\boldmath$\uiota$}}

\def\beq{\begin{equation}}
\def\eeq{\end{equation}}
\def\beqa{\begin{eqnarray}}
\def\eeqa{\end{eqnarray}}
\def\beqan{\begin{eqnarray*}}
\def\eeqan{\end{eqnarray*}}

\def\bc{\begin{center}}
\def\ec{\end{center}}
\def\btable{\begin{table}[htbp]}
\def\etable{\end{table}}
\def\bfig{\begin{figure}[htbp]}
\def\efig{\end{figure}}

\def\bi{\begin{itemize}}
\def\ei{\end{itemize}}
\begin{document}
\doublespace
\noindent{\huge\bf  A Statistical Approach to Ecological Modeling by a New Similarity Index
 }\\
 \
 
\noindent Srijan Chattopadhyay,$^{1,2}$ and Swapnaneel Bhattacharyya$^{1,3}$  \\
\
\noindent{$^1$ Indian Statistical Institute, 203 B.T. Road, Kolkata-700108, India}\\
\
\noindent{$^2$ \href{mailto:srijanchatterjee123456789@gmail.com}{srijanchatterjee123456789@gmail.com}}\\
\
\noindent{$^3$
\href{mailto:swapnaneelbhattacharyya@gmail.com}{swapnaneelbhattacharyya@gmail.com}

Abstract: Similarity index is an important scientific tool frequently used to determine whether different pairs of entities are similar with respect to some prefixed characteristics. Some standard measures of similarity index include Jaccard index, Sørensen-Dice index, and Simpson's index. Recently, a better index ($\hat{\alpha}$) for the co-occurrence and/or similarity has been developed, and this measure really outperforms and gives theoretically supported reasonable predictions. However, the measure $\hat{\alpha}$ is not data dependent. In this article we propose a new measure of similarity which depends strongly on the data before introducing randomness in prevalence. Then, we propose a new method of randomization which changes the whole pattern of results. Before randomization our measure is similar to the Jaccard index, while after randomization it is close to $\hat{\alpha}$. We consider the popular ecological dataset from the Tuscan Archipelago, Italy; and compare the performance of the proposed index to other measures. Since our proposed index is data dependent, it has some interesting properties which we illustrate in this article through numerical studies. 

\section{Introduction}
Similarity or Co-occurrence analysis is an interesting  research area for a long time. It plays an important role in many disciplines including Ecology (Gotelli and McCabe 2002; Dornelas et al. 2014; Richer de Forges et al. 2000; Blowes et al. 2019; Oluyinka Christopher 2020, Plata et al. 2015), Medical Science (Ocampo et al., 2014), Genetics (Gianola et al., 2012), Neuroscience (Crossley et al., 2013), Economics (Kellman and Schroder, 1983), Geology (Nazari et al., 2016) etc. Researchers frequently use similarity indices and co-occurrence analyses to determine the distributional similarities between two populations in the same time period, or for the same population in different time periods. In medical sciences, it is often important to determine if two drugs work similarly once administered on a comparable group of patients. In the genetic studies it is of interest to investigate if a given pair of genes function similarly in the development of some biological trait(s) or disease(s).  

In literature, three most commonly used similarity indices are Jaccard index, Sørensen-Dice Index and Simpson's index. We define these three indices below for completeness.
\begin{itemize}
    \item \textbf{Jaccard Index}: This index was proposed by Paul Jaccard (1912), and it measures the similarity between finite sample sets. It is defined as the size of the intersection divided by the size of the union of the sample sets, $J(A, B) = \frac{|A \cap  B|}{|A \cup B|}$. So, by definition, $0 \leq J \leq 1.$

    \item \textbf{Sørensen-Dice Index:} This index was proposed by Sørensen (1948), and is defined as $\frac{2 |X \cap Y|}{|X| + |Y|}$, for two given finite sets $X$ and $Y$. 

    \item \textbf{Simpson's Index:} Simpson’s Diversity Index is a statistic to measure the diversity of species in a community (Simpson, 1949). In a population with $N$ entities if there are $R$ communities and $n_i$ denotes the number of entities belonging to the i-th community, then the Simpson's Index is defined as $\frac{\sum\limits_{i=1}^R{n_i(n_i-1)}}{N(N-1)}$. This can be interpreted as the probability of two entities coming from the same community when the entities are randomly selected from the population. 
    \end{itemize}
    All these indices interpret smaller values (closer to 0) as a highly negative association and a value near one as a highly positive one. More recently, a new measure of similarity  $(\hat{\alpha})$ has been proposed in Mainali et al. (2022). These authors claim that the three popularly used metrics defined above (and numerous other related measures) suffer from major statistical flaws which consequently introduce bias. They provide an affinity model based on some unknown parameters (probabilities) $p_1$ and $p_2$, and propose a theoretical measure $\alpha = log \left(\frac{p_1}{1-p_1}/\frac{p_2}{1-p_2}\right)$, and the MLE of $\alpha$ is their proposed similarity index. We reproduce the following propositions from Mainali et al. (2022) for a better understanding of their proposed measure. 
    \begin{itemize}
        \item \textbf{Proposition 1:} \textit{ (Null hypothesis distribution) If the numbers $m_A$, 
$m_B$, and $N$ are fixed, and $m_A$ balls colored $A$ are placed in boxes (with at most one ball to a box) in any way whatever and given the placement of the $A$-colored balls, the $m_B$ balls colored $B$ are placed  equiprobably in the $N$ boxes, then the probability distribution of $X$, the count of boxes containing balls of both colors is hypergeometric $(N, m_A, m_B)$}.

   \item \textbf{Proposition 2:} \textit{(Affinity model distribution) Suppose that the numbers $m_A, m_B$ and $N$ are fixed, and $m_A$ balls colored $A$ are placed in boxes (with at most one ball in each box) in any way whatever and given the placement of the $A$-colored balls balls colored $B$ are placed independently in boxes with probability $p_1$, for boxes containing $A$-colored balls, and with probability $p_2$ for boxes not containing  $A$-colored balls. Then, conditionally given that a total of $m_B$ balls are placed in boxes, either conditionally given the configuration of the 
$A$-colored balls or unconditionally, the distribution of $X$, the count  of co-occurrences is an extended hypergeometric distribution with parameters $N, m_A, m_B$, and $\alpha = log (\frac{p_1}{1-p_1}/\frac{p_2}{1-p_2})$}.
    \end{itemize}

    Hence, \begin{equation}
    P(X=k) = \frac{ {m_A \choose k} {N-m_A \choose m_B-k}e^{\alpha k} }{\sum\limits_{j=s}^{t}{m_A \choose j} {N-m_A \choose m_B-j}e^{\alpha j}},
    \end{equation}
    
     for $s=max(m_A + m_B - N,0) \leq k \leq min (m_A, m_B)=t$. Here $X$ is the count of co-occurrences. This is an  extended hypergeometric or non-central hypergeometric distribution. So, for observed $X=k$ with fixed $(N,m_A,m_B)$ Mainali et al. (2022) estimated $\alpha$ by its maximum likelihood estimate $\hat{\alpha}$, and conclude that it is the unique solution to the equation given below when a finite solution exists: 
     \begin{equation}
         X = \frac{\sum\limits_{j=s}^{t} j{m_A \choose j} {N-m_A \choose m_B-j}e^{\alpha j}}{\sum\limits_{j=s}^{t}{m_A \choose j} {N-m_A \choose m_B-j}e^{\alpha j}}.
     \end{equation}

     For $X=t$, Mainali et al. (2022) defined $\hat{\alpha}$ to be $-\infty$, and it is $+\infty$ for $X=s$. 
     
     The authors mentioned that the most interesting feature of this measure is that its distribution hardly depends on the prevalence i.e. on the $(N,m_A,m_B)$, whereas the mostly used three indices do depend on the prevalence. Motivated by this measure, we propose a new measure of similarity index which also does not directly depend on the prevalence but it is data dependent. In other words, for the same values of the prevalence different datasets provide different values of the index. Since our proposed measure is data dependent, it has some additional benefits which we will discuss in the later sections. 

     We analyze a biogeographical data to measure the diversity of ecological communities across regions, or compositional change of a single community over different time periods. This dataset has been analyzed earlier in the literature using Jaccard index and $\hat{\alpha}$, but the results were completely in the opposite directions. We show that our measure is quite sensitive in the sense that once we introduce a stochastic component it gives results similar to that for $\hat{\alpha}$. However, without the stochastic part we get results similar to the Jaccard index. 

     The rest of this article is organized as follows. In Section 2, we propose our data dependent similarity index, and discuss some of its features. In Section 3, we summarize the results from the data application before introducing the stochastic component in the model. In Section 4, we introduce the randomization and re-analyze the dataset, and summarize the results. Finally, some concluding remarks are given in Section 5. 
    
\section{Model and Method} 
We describe our model in the context of the ecological modeling described in Section 1. Consider
 $k$ species, i.e. $S_1, \ldots, S_k$; and we collect data from $n$ islands $I_1, \ldots, I_n$ over $l$ time periods $t_1, \ldots, t_l$. We assume that for a fixed time period, say $r$, the species $S_i$ can occur in any of the $n$ islands independently with the same probability $p_{ir}$. Now, for each time period, we construct a $k \times n$ matrix $S^{r}$, so that the $(i,j)$-th element of $S^{r}$ which we denote by $S_{ij}^{r}$ is 1 if the $i$-th species is present in the $j$-th island at time period $r$; and 0, otherwise. 

 For estimating $p_{ir}$ by the maximum likelihood estimation, we need to average the rows of the matrix $S^{r}$. For some $i$, if the corresponding row sum is zero, we take the row sum as 1 and compute the average as MLE. On the other hand, if the row sum is $n$, then we consider it as $n-1$, and find the corresponding MLE. This manipulation is needed to ensure that the MLE $\hat{p}_{ir}$ lies in $(0,1)$.

 Next, we transform the matrix $S^{r}$ for making its elements continuous. We consider the logit and probit type of transformation since the elements of $S^{r}$  are binary. A logit type transformation results in the following matrix $T^{r}$ with the $(i,j)$-th element as $T^r_{ij}$ given below:
$$
T^r_{ij} = \begin{cases}
                log\left(\frac{\hat{p}_{ir}}{1 - \hat{p}_{ir}}\right), & \text{if $S^r_{ij} = 1,\; \hat{p}_{ir} \geq \frac{1}{2}$},\\
                log\left(\frac{1-\hat{p}_{ir}}{ \hat{p}_{ir}}\right), & \text{if $S^r_{ij} = 0, \; \hat{p}_{ir} \geq \frac{1}{2}$},\\
                log\left(\frac{\hat{p}_{ir}}{1 - \hat{p}_{ir}}\right), & \text{if $S^r_{ij} = 0, \;  \hat{p}_{ir} < \frac{1}{2}$},\\
                log\left(\frac{1-\hat{p}_{ir}}{ \hat{p}_{ir}}\right), & \text{if $S^r_{ij} = 1, \; \hat{p}_{ir} < \frac{1}{2}$}.\\
            \end{cases}$$

   For a probit type of transformation, we get the following:
$$
T^r_{ij} = \begin{cases}
                \Phi^{-1}(\hat{p}_{ir}), & \text{if $S^r_{ij} = 1, \; \hat{p}_{ir} \geq \frac{1}{2}$},\\
                \Phi^{-1}(1-\hat{p}_{ir}), & \text{if $S^r_{ij} = 0, \; \hat{p}_{ir} \geq \frac{1}{2}$},\\
                \Phi^{-1}(\hat{p}_{ir}), & \text{if $S^r_{ij} = 0, \; \hat{p}_{ir} < \frac{1}{2}$},\\
                \Phi^{-1}(1-\hat{p}_{ir}), & \text{if $S^r_{ij} = 1, \; \hat{p}_{ir} < \frac{1}{2}$},\\
            \end{cases}$$  

where $\Phi$ denotes the CDF of $N(0,1)$.

\subsection{Principal Component Analysis and ``score" matrix} 

We note that corresponding to each time period we get a matrix of Continuous data entries by the logit/probit transformation. So we get the matrices $T^1, \ldots, T^l$, where $T^r$ corresponds to the $r$-th time period. Note that even for moderate values of $l$ and $n$, large $k$ will result in a large number of parameters to handle. Therefore, we implement principal component analysis for reducing the dimension and also for the ease of computation (Jolliffe and Cadima, 2016).

 We define a new matrix $X$ as follows:
\begin{align}
    X &= \begin{bmatrix}
           {T^1}^T \\
           {T^2}^T \\
           \vdots \\
           {T^l}^T
         \end{bmatrix},
  \end{align}
so that $X$ has $k$ columns and $n \times l$ rows. Now we center the matrix $X$, and define matrix $Y$ as follows: 
\begin{align}
    \begin{bmatrix}
           Y_{1,i} \\
           Y_{2,i} \\
           \vdots \\
           Y_{nl,i}
         \end{bmatrix} &= \begin{bmatrix}
           X_{1,i} - \Bar{X}_i \\
           X_{2,i} - \Bar{X}_i \\
           \vdots \\
           X_{nl,i} - \Bar{X}_i
         \end{bmatrix},
  \end{align}
  where $\Bar{X}_i = \frac{X_{1,i} + ... + X_{nl,i}}{nl}.$
  
  Now we focus to find the eigen vectors of $Y^TY$ which of dimension $k \times k$. Usually, in the ecological studies $n \times l$ is much smaller than $k$. We note that,
  \begin{center}
      $YY^T \Vec{v} = \lambda \Vec{v}, $\\
      $Y^TY ( Y^T\Vec{v}) = \lambda (Y^T\Vec{v}).$ 
  \end{center}
  So if we can find the eigen vectors of $YY^T$, then by right multiplying by $Y^T$ we can get eigen vectors of $Y^TY$. Now let the eigen values of $YY^T$ be $\lambda_1 \geq \lambda_2 \geq ... \geq \lambda_n \geq 0$.  By considering only the ``important" eigen values based on a prefixed threshold (in our application, we consider eigen values $\geq 10^{-5})$ we select a subset of eigenvalues. Then by considering the corresponding eigen vectors  and by adjoining them column-wise into a matrix we get a matrix, say $P_1$. Define $P=Y^TP_1$. Clearly, $P$ is the set of eigen vectors of $Y^TY$ corresponding to the ``important" eigen values. Then, we normalize each column and call the new matrix $Q$. Thus, we get an orthonormal basis for the row space of $Y$. Next, we express the rows of $Y$ in terms of this orthonormal basis, and get a score  vector for each of the rows of $Y$ i.e. for each of the columns of the matrix $S^r$. Thus, the score is $YQ$, and now we are in a space whose dimension is smaller than $ \min \{k,n \times l\}$. The principal component analysis helps us to achieve this.

  \subsection{Linear Discriminant Analysis to Score matrix}

  Note that our objective is to characterize or separate different islands with respect to the species, and therefore we consider the linear combination of the columns of the score matrix and apply Fisher's linear discriminant analysis similar to Demir and Ozmehmet (2005). Let $R$ be the score matrix from Section 2.1. Define the matrices $U_1, \ldots, U_l$ such that $U_1 = R[1:n,], U_2 = R[n+1:2n,], \ldots, U_i = R[(i-1)n+1:in,] $ etc. Next, we define $Y_1, \ldots, Y_l$ in the similar way we defined $Y$ from $X$ in Section 2.1, i.e. by centering the data. Then, define $m_1, \ldots, m_l$ such that $m_i$ is the vector consisting of the mean of the elements in each column of $U_i$. Then construct matrix $Z$ as follows: 
  \begin{align}
    Z &= \begin{bmatrix}
           {U_1} \\
           {U_2} \\
           \vdots \\
           {U_l}
         \end{bmatrix}.
  \end{align}
We note that if there are $C$ classes, $i$-th class with mean $\mu_i$ and the common covariance matrix $\Sigma$, then the scatter between class variability can be defined by the sample covariance of the class means as $\Sigma_b = \frac{1}{C}\sum\limits_{i=1}^C (\mu_i - \mu)(\mu_i - \mu)^T$,  where $\mu$  is the mean of the class means. The class separation in a direction $\Vec{x}$ is given by 
$S = \frac{\Vec{x}^T\Sigma_b\Vec{x}}{\Vec{x}^T\Sigma\Vec{x}}$.

So, we have to find the direction which maximizes $S$ because in that direction the variation within the same class ($W$) is minimized and variation between different classes ($B$) is maximized. In our case, $W = c Z^TZ$, where $c$ is a constant which represents $\Sigma_b$, and $B = d \hat{\Sigma}$, where $\hat{\Sigma}$ is covariance matrix of $m_1,\ldots,m_l$, and $d$ is a constant. Thus, we have to find the following:
\begin{center}
    $\underset{\Vec{x} \in \mathbb{R}^s \setminus {\Vec{0}}}{\text{arg max : }} 
 \frac{\Vec{x}^T W\Vec{x}}{\Vec{x}^T B \Vec{x}}.$
\end{center}

We use the following result for finding this. The proof of the result is given in the appendix. 

\begin{theorem}
   Let $W$ be Positive Definite. Then, 
    $\underset{\Vec{x} \not= \Vec{0}}{\text{max}} \frac{\Vec{x}^T B\Vec{x}}{\Vec{x}^T W \Vec{x}} = \lambda_1(W^{-1}B)$, 

where $\lambda_1(W^{-1}B)$ is the largest eigen value of $W^{-1}B$, and the maximum is attained at $\Vec{x}_0$ if and only if $\Vec{x}_0$ is an eigen vector of $W^{-1}B$ corresponding to $\lambda_1(W^{-1}B)$. 
\end{theorem} 

In practice, we may come across the following issues:
\begin{itemize}
    \item From the proof it is evident that the eigen values of $B^{-1}W$ must be real. But the algorithmic executed in statistical softwares (e.g. R) does not always give real eigen values as output. In that case, following the proof, we do the Cholesky decomposition of $B$ and ask to return the corresponding eigen value. So using Cholesky decomposition we get the following:
    \begin{center}
        $W = M^TM$, \medskip \\
        $(M^{-1})^T B M^{-1} \Vec{y}_0 = \lambda_1 \Vec{y}_0$,\medskip \\
        $M^{-1}(M^{-1})^T B (M^{-1} \Vec{y}_0) = \lambda_1 (M^{-1}\Vec{y}_0)$,\medskip \\
        $W^{-1}B \Vec{x}_0 = \lambda_1 \Vec{x}_0,$
    \end{center}
    where $\Vec{x}_0 = M^{-1}\Vec{y}_0$. Instead, we consider $\Vec{z}_0 = M\Vec{y}_0 = M^2\Vec{x}_0$ and look at the scores with respect to $\Vec{z}_0$.

    \item Also $W$ may not always be invertible. Then we follow the following technique,  we take some small positive $\lambda$ and add $\lambda I$ to $W$ to make it invertible.  
     \end{itemize}

After this, we get a matrix consisting of eigen vectors of $W^{-1}B$ as columns. Now, we want to express $R$ in terms of $M^{2}S$, and we introduce a new matrix $Scores_{new}$ as: $Scores_{new} = R (S^T_1)^{-1}$, where $S_1 = M^2S$. Hence, the first column of $Scores_{new}$ will be the scores with respect to $\Vec{z}_0$. Let $N = scores_{new}[,1]$. Then, $N$ is a vector of dimension $n \times l$.
Our similarity index (SI) for $i$-th island will be defined as: $SI_i=sd(N_i,N_{i+n},..,N_{i+n(l-1)})$, where sd stands for the standard deviation. So the possible range of this index is $(-\infty, 1]$, and it is closer to 1 indicates a better association/similarity, and the more negative indicates more dissimilarity.
    
\section{Data Applications}
We analyze two different ecological datasets which have already analyzed and reported in the literature. We re-analyze these datasets by our newly proposed similarity index, and compare our findings with the ones already reported. 

\subsection{Application 1: Bio-geography: similarity between different islands}

We apply our proposed similarity index to a bio-geographical dataset already analyzed in Mainali et al. (2022). These authors computed Jaccard index and $\hat{\alpha}$, and it was shown that these two indices provides different results. We show our proposed non-random measure provides the same pattern as of Jaccard index, and it happens because of the non-stochastic behaviour of our index.

Beta-diversity is a measure that indicates the diversity of ecological communities across space or compositional change in a single community over different time periods. So, a low beta-diversity signifies more similarities or less significant changes over space or time. The dataset consists of 16 islands of the Tuscan Archipelago, Italy,  for the time periods 1830 to 1950, and 1951 to 2015. The latitudes of the islands are all in the range of 42.25 to 43.43, and the longitudes are in the range of 9.82 to 11.32. Hence, the islands are in the very comparable environments and therefore it is reasonable to assume that  the probability of observing a species is the same for all islands in the same time period. The dataset contains a total of 1842 species for each of which we have data for 16 islands in two different aforementioned time periods. For details on the data description, we refer to Chiarucci et al. (2017). 

Mainali et al. (2022) analyzed this dataset, and reported their index $\hat{\alpha}$, and the Jaccard index, and noted that the results are quite in opposite direction. We analyze this dataset using our proposed similarity index. For each of the 16 islands we compute the proposed similarity index for measuring the similarity of that particular island for two different time periods. We use both the logit transformation and the probit transformation, and compute two similarity indices for each island which we denote by  $\hat{\beta_L}$ and $\hat{\beta_N}$, respectively. 

As reported in Mainali et al. (2022), $\hat{\alpha}$ shows that larger islands are less stable (in terms of the species) than the smaller one, but the Jaccard index shows the larger islands are more stable. In Figure 1, we plot the log transformed area and the computed similarity index for each of the 16 islands, and fit a simple linear regression. Plots for $\hat{\beta_L}$ and $\hat{\beta_N}$ show increasing trends and illustrate that for the larger islands more similarity observed over two time periods.  

 Similarly, in Figure 2, we see that as the distance to the nearest bigger island increases $\hat{\beta}_N$ and $\hat{\beta}_L$ increases. 
\begin{figure}[h!]
    \centering
    \includegraphics[scale = 0.6]{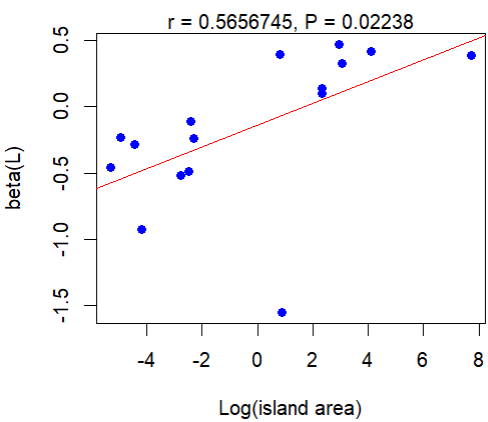}
    \includegraphics[scale = 0.6]{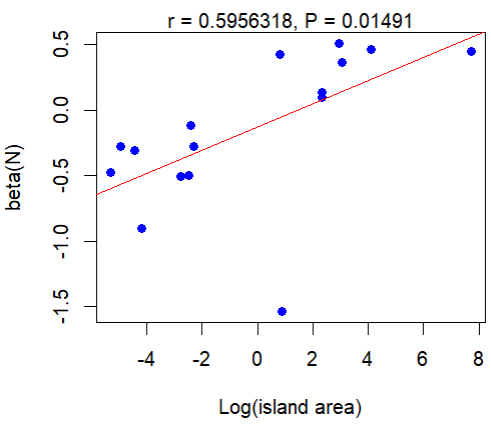}
    \caption{Regression plot for beta(L) and beta(N) with island area}
    \label{fig:my_label}
\end{figure}
\begin{figure}[h!]
    \centering
    \includegraphics[scale = 0.6]{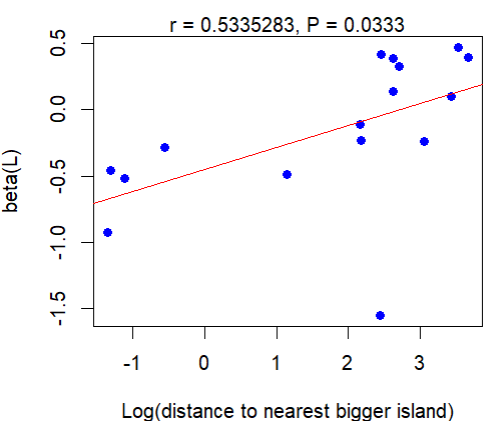}
    \includegraphics[scale = 0.6]{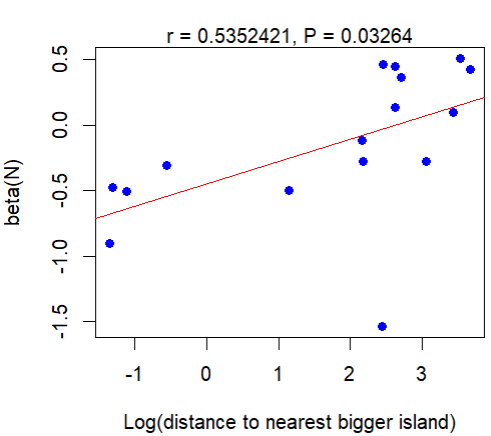}
    \caption{Regression plot for beta(L) and beta(N) with distance to nearest bigger island}
    \label{fig:my_label}
\end{figure}
\begin{figure}[h!]
    \centering
    \includegraphics[scale = 0.6]{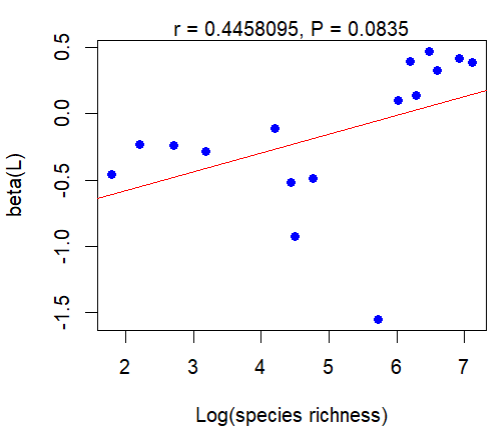}
    \includegraphics[scale = 0.6]{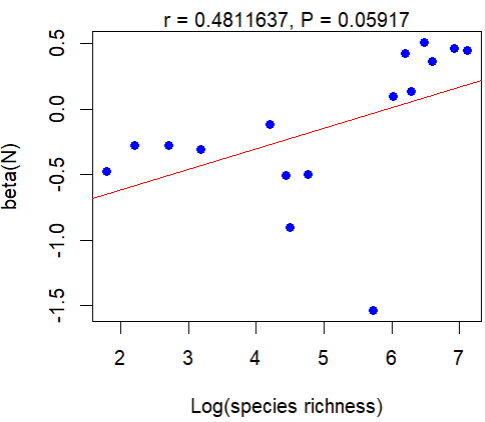}
    \caption{Regression plot for beta(L) and beta(N) with species richness}
    \label{fig:my_label}
\end{figure}

In Figure 3, we see that there is an increasing trend for our proposed index for species richness. This finding is already reported in Kraft et al. (2011) who concluded that beta-diversity decreases with species richness and latitude just as a consequence of the decreasing size of the species pool. All our findings for this dataset are in the line of Jaccard index, but are in opposite direction to that of $\hat{\alpha}$.  

\subsection{Application 2: Time similarity between islands and space similarity between species}

Here, we consider the beta-diversity over time. We consider a dataset previously analyzed by Tulloch et al. (2018). The dataset contains 88 bird species from 96 islands for three different years 2011, 2012, and 2013. We first want to see the time similarities between various species, and then will see the similarities between different bird species throughout the years.  In this application, we define time similarity for three years as $sd(S_{k,1}, S_{k,2}, S_{k,3})$, with $S_{k,j}$ as the score obtained in the $j$-th time period for the $k$-th island.  For the index between any two time periods we use the same definition as before. We compute $\hat{\beta_N}$ and $\hat{\beta_L}$ similar to Section 3.1, and  show a scatter plot  and the QQ plot in Figure 4. We note that both these plots indicate that these two indices are very much close to each other in numerical values. The joint similarity (over three years period) of the islands are shown in Figure 5. The plots of $\hat{\beta}_L$ and $\hat{\beta}_N$ look very similar.

\begin{figure}[h!]
    \centering
    \includegraphics[scale = 0.6]{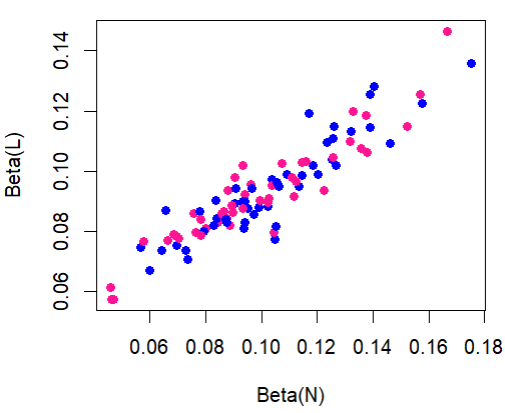}
    \includegraphics[scale = 0.6]{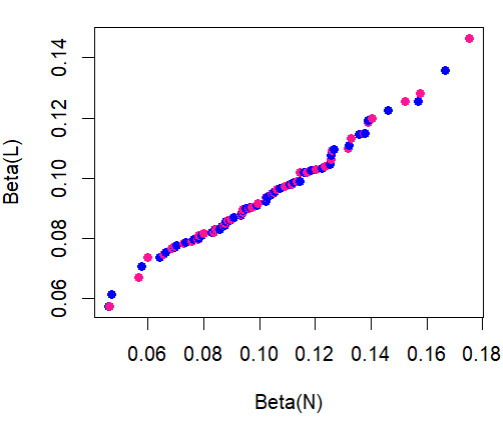}
    \caption{Scatter and QQ plot of $\hat{\beta_L}$ and $\hat{\beta_N}$ }
    \label{fig:my_label}
\end{figure}
\begin{figure}[h!]
    \centering
    \includegraphics[scale = 0.6]{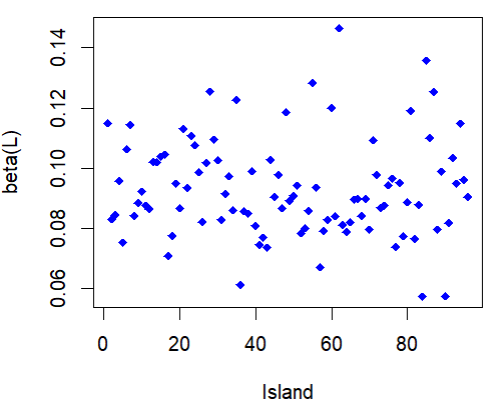}
    \includegraphics[scale = 0.6]{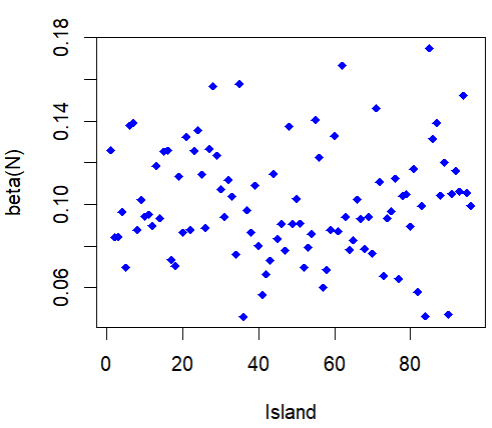}
    \caption{Plot of the joint similarity $\hat{\beta_L}$ and $\hat{\beta_N}$ of the islands}
    \label{fig:my_label}
\end{figure}
Next, we move to the analysis of similarity among different bird species. We project the whole similarity matrix into several heatmaps time-wise and also as a whole for better understanding. In each of the heatmaps we see there is a white off-diagonal line which indicates a value close to 1 because it is the self-similarity. The white portions indicate stronger similarity than the blue portions. As the score is bounded by 1, hence there is no red box in the figures according to the color scale. Based on figures 6-9, we can infer the following:
\begin{itemize}
    \item In 2013, the species Golden Whistler, Brown Quail, Varied Sittella, and Southern Whiteface were most similar as we can see from the central white part of the heat map for 2013.
    \item In 2012, the species Little Corella, Golden Whistler, Apostlebird, Brown Quail, and Sittella were very similar as we can see from the lower white part of the heat map for 2012.
    \item In 2011, the species Little Corella, Apostlebird, Little Raven behaved similarly as evident from the heat map for 2011.
    \item The overall heat map for three years as shown in Figure 6 shows that there are more while spots than blue spots. This indicates that more group of birds were similar than the groups who behaved very differently. 
\end{itemize}
\begin{figure}[h!]
    \centering
    \includegraphics[scale = 0.4]{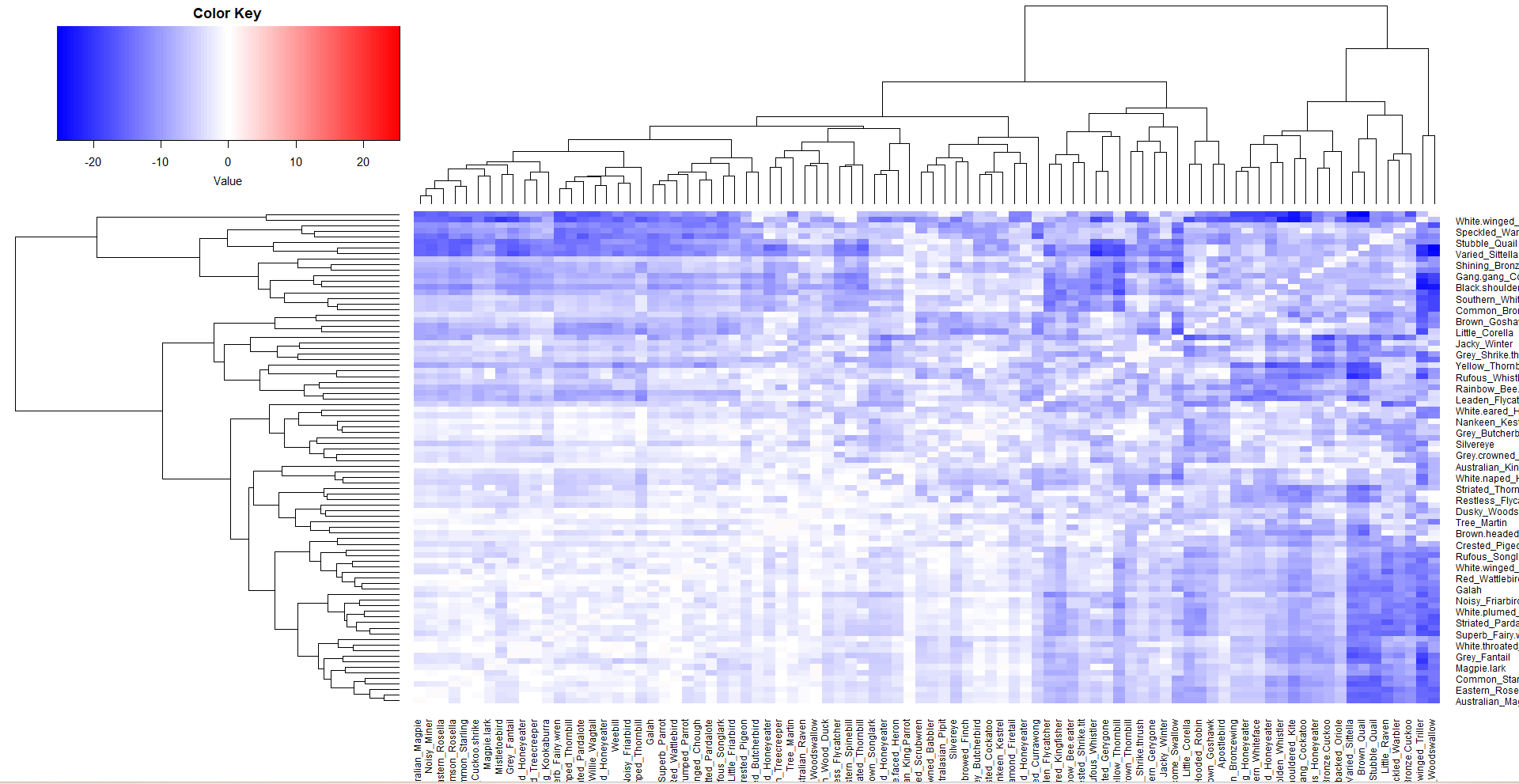}
    \caption{Heat map of the similarity of the species for all three years}
    \label{fig:my_label}
\end{figure}
\begin{figure}[h!]
    \centering
    \includegraphics[scale = 0.4]{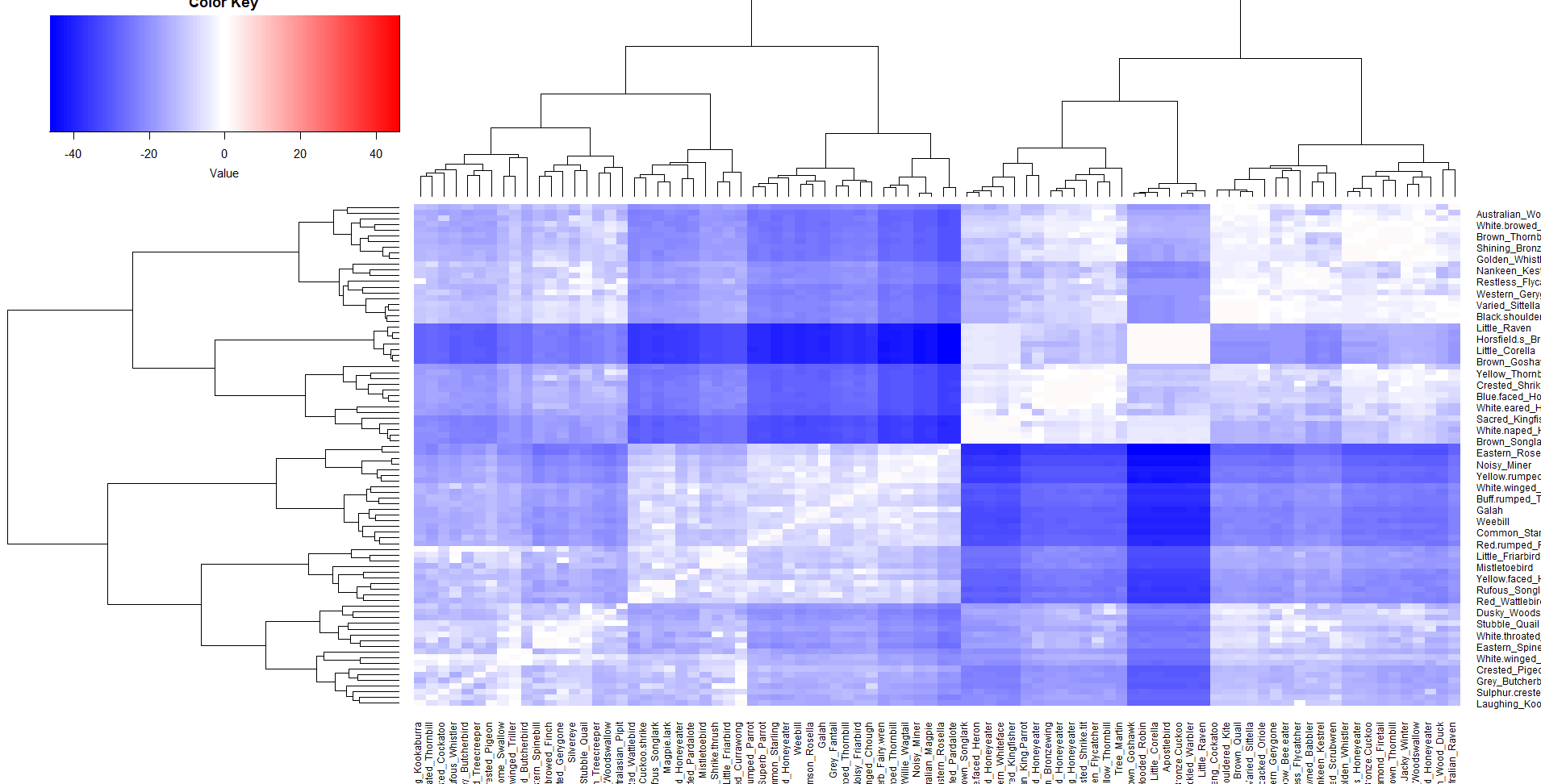}
    \caption{Heat map of the similarity of the species for 2011}
    \label{fig:my_label}
\end{figure}
\begin{figure}[h!]
    \centering
    \includegraphics[scale = 0.4]{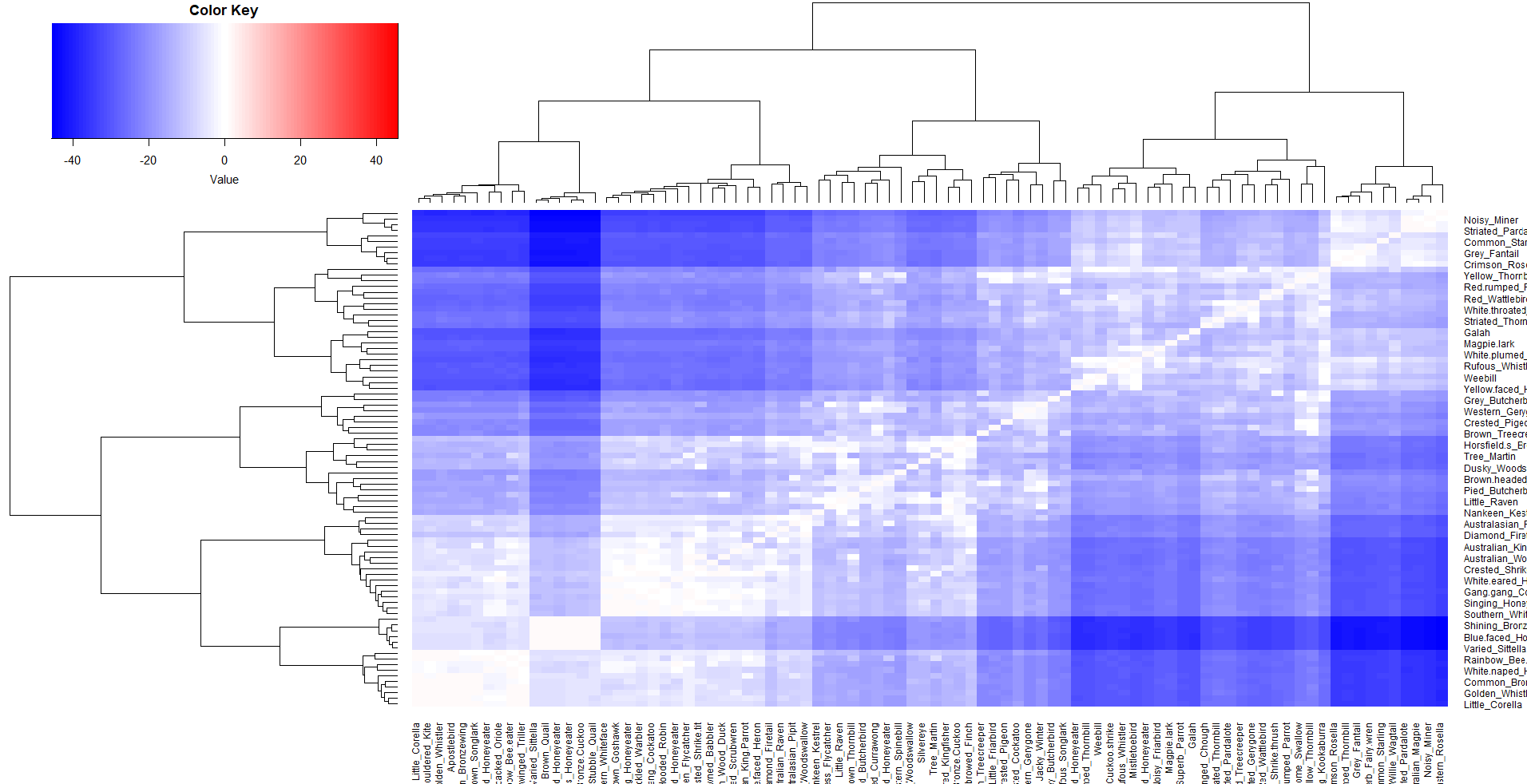}
    \caption{Heat map of the similarity of the species for 2012}
    \label{fig:my_label}
\end{figure}
\begin{figure}[h!]
    \centering
    \includegraphics[scale = 0.4]{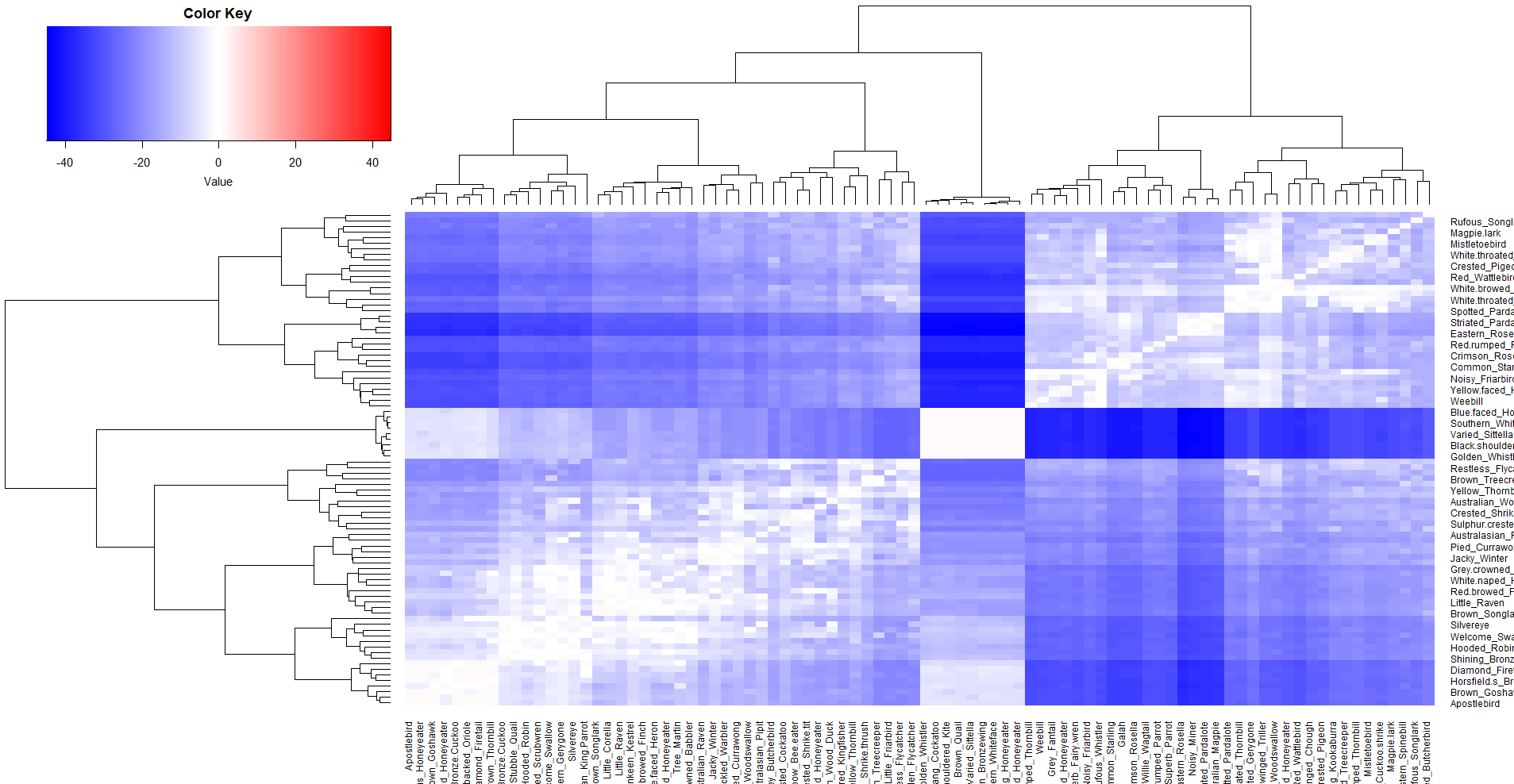}
    \caption{Heat map of the similarity of the species for 2013}
    \label{fig:my_label}
\end{figure}


\vspace{10cm}
\subsection{Randomization}
In the analysis reported in Section 3.1 and 3.2, for different time periods we keep the marginal distributions of each species fixed. In that case, our proposed index highly depends on the prevalence since there is no stochastic component in the model. For that we introduce the method of randomization in our index as follows.  The initial distribution of $X_0$ which denotes the presence of each species in each time period is given by $\mu = \begin{bmatrix}
    p & 1-p
\end{bmatrix},$ where $p = 1-\hat{p}_{i,r}$, (if it is done for
$i$-th species and for $r$-th time period).
We change the entries 0 or 1 randomly with the following probabilities:  
\begin{center}
    $P(X_1 = 0| x_0 = 0) = a,$\\
    $P(X_1 = 1| x_0 = 0) = 1-a,$\\
    $P(X_1 = 0| x_0 = 1) = b,$\\
    $P(X_1 = 1| x_0 = 1) = 1-b,$
\end{center}where, $X_1$ denotes the entries after the randomization. So the marginal distribution of $X_1$ is given as follows:
\begin{center}
    $X_1 \sim \begin{bmatrix}
        p & 1-p
    \end{bmatrix} \begin{bmatrix}
        a & 1-a \\
        b & 1-b
    \end{bmatrix}. $
\end{center}
Now, this transformation will have the following properties: 
\begin{itemize}
    \item The marginal distribution of $X_0$ and $X_1$ should be the same. 
    \item The Kullback-Leibler Distance between $X_0$ and $X_1$ should be minimum.
\end{itemize}
To satisfy the first condition, we need to select $a,b$ so that the following equation holds:
 
\begin{center}
    $\begin{bmatrix}
        p & 1-p
    \end{bmatrix} = \begin{bmatrix}
        p & 1-p
    \end{bmatrix} \begin{bmatrix}
        a & 1-a \\
        b & 1-b
    \end{bmatrix}.$
\end{center}
So we have the following equation : 
\begin{equation}
    ap + b(1-p) = p,
    \end{equation}
which gives the following relation between $a,b$:
\begin{equation}
    b = \frac{p}{1-p}(1-a), \;\; 0<p<1, 0<a<1.
\end{equation}
Now we choose $a$ so that the Kullback-Leiblar Divergence between $X_0$ and $X_1$ is minimal. We at first calculate the Kullback-Leiblar Distance between $X_0$ and $X_1$. The joint distribution of $(X_0,X_1)$ is given by:
\begin{center}
    $p(0,1) = b(1-p)$,\medskip\\
    $p(1,0) = (1-a)p$,\medskip\\
    $p(0,0) = ap$,\medskip\\
    $p(1,1) = (1-b)(1-p).$
\end{center}

Thus, the Kullback-Leibler Divergence $D$ is given by: 
\begin{center}
    $D = p(0,1) log(\frac{p(0,1)}{p(0)p(1)}) + p(1,0) log(\frac{p(1,0)}{p(1)p(0)}) + p(0,0) log(\frac{p(0,0)}{p(0)p(0)}) + p(1,1) log(\frac{p(1,1)}{p(1)p(1)}).$
\end{center}

Hence,
\begin{align*}
    D=b(1-p)log\left(\frac{b(1-p)}{p(1-p)}\right)+(1-a)plog\left(\frac{(1-a)p}{p(1-p)}\right)+\\aplog\left(\frac{ap}{p^2}\right)+(1-b)(1-p)log\left(\frac{(1-b)(1-p)}{(1-p)^2}\right)\\
    = b(1-p)log\left(\frac{b}{p}\right)+(1-a)plog\left(\frac{(1-a)}{(1-p)}\right)+aplog\left(\frac{a}{p}\right)+\\(1-b)(1-p)log\left(\frac{(1-b)}{(1-p)}\right).
\end{align*}
Since from (6) it is noted that optimum $b$ is a function of $a$, hence this expression is a function of $a$. So we get,
\begin{equation}
    D(a) = 2 (1-a)p log\left(\frac{1-a}{1-p}\right) + a p log\left(\frac{a}{p}\right) + (1 - 2p + ap) log\left(\frac{1 - 2p + ap}{(1-p)^2}\right).
\end{equation}
\begin{equation}
   D'(a)=-2p\left(log\left(\frac{a-1}{p-1}\right)+1\right)+p\left(log\left(\frac{a}{p}\right)+1\right
   )+p\left(log\left(\frac{p(a-2)+1}{(p-1)^2}\right)+1\right).
\end{equation}
$D'(a) = 0$ yields the solution $a = p$, since $D(p) = 0$, and by Gibbs' Inequality, $D(a) \geq 0$, 
  $\forall a \in (0,1)$, and so \textbf{$p = \underset{a \in (0,1)}{\text{arg min}} D(a)$}.
  
We apply this randomization $M$ times, for large $M$. Then for each step we get a score vector and store them in a matrix row-wise. Then we normalize each row (i.e. scale down the norm of every row to 1) and take the column-wise mean to get an average score vector of all the randomized datasets. We call this vector $N$, a vector of dimension $nl \times 1$.\\
So our similarity index (after randomisation) for $i$-th island will be $1 - \text{standard deviation of}$ $N_i,N_{i+n},..,N_{i+n(l-1)}$.
 The range of this index is $(-\infty, 1]$, and the it is close to 1 indicates more association and the more negative indicates more  dissimilarity, as earlier.

\subsubsection{Results after Randomization}
We note that the main goal of randomization is to vary through all the prevalence sets by keeping the marginal of each species in each time period fixed. So, we perform three sets of randomisations with $N=1000, 2000$, and $5000$, respectively. We notice consistency in our results. That is, in each such randomization set we get very similar results. And surprisingly we reanalyze the island dataset using the randomization method in our proposed index, and the results are summarized in Figures 10-12. As we see the whole trend goes in favor of $\hat{\alpha}$, i.e. larger islands are less stable. Here, we only show the results using $\hat{\beta}_L$, but $\hat{\beta}_N$ yields very similar results. Thus, after randomization $\hat{\alpha}$ matches with $\hat{\beta}_L$ in terms of the inference, and this illustrates the flexibility of our proposed index.

\begin{figure}[h!]
    \centering
    \includegraphics[scale = 0.6]{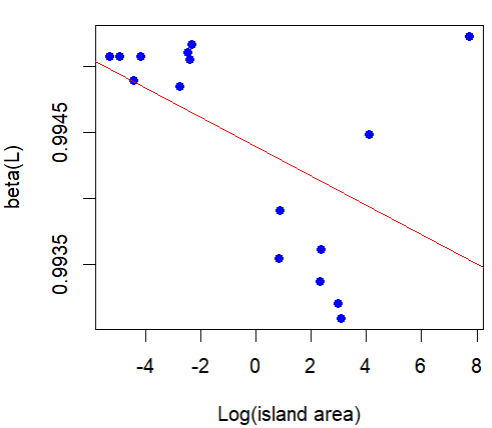}
    \includegraphics[scale = 0.6]{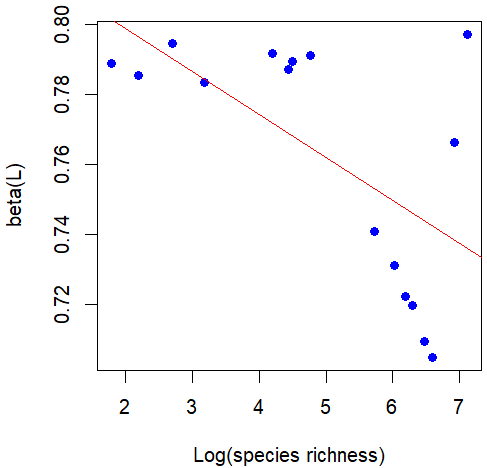}
    \caption{Results after first set of (1k) randomizations}
    \label{fig:my_label}
\end{figure}
\begin{figure}[h!]
    \centering
    \includegraphics[scale=0.6]{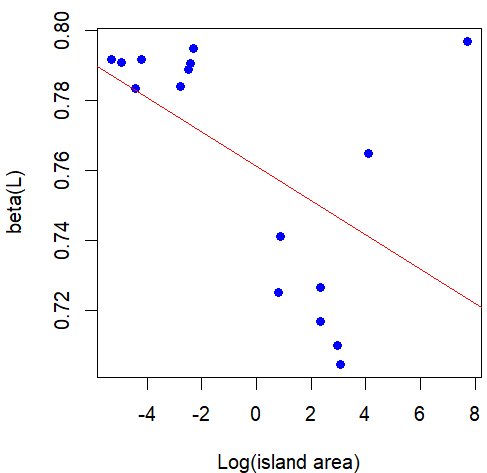}
    \includegraphics[scale=0.6]{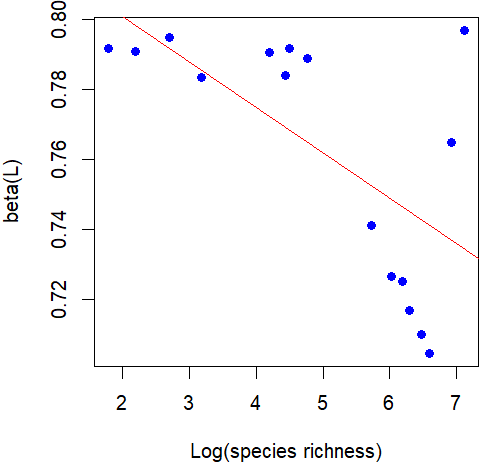}
    \caption{Results after second set of (2k) randomizations}
    \label{fig:my_label}
\end{figure}
\begin{figure}[h!]
    \centering
    \includegraphics[scale=0.6]{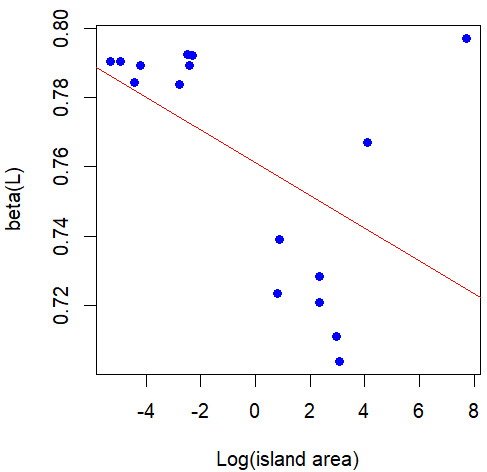}
    \includegraphics[scale=0.6]{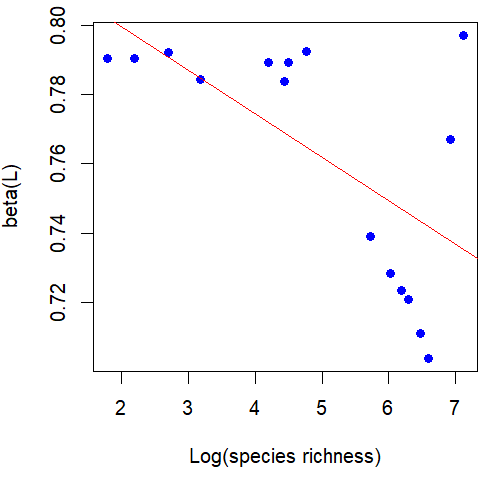}
    \caption{Results after third set of (5k) randomizations}
    \label{fig:my_label}
\end{figure}

\section{Conclusion}
In many applications similarity indices are used for data-based scientific inferences. In this article, we have developed a new similarity index and have illustrated the usefulness of it in two ecological datasets. Our proposed index is flexible in the sense that it is data-dependent and also can be adjusted by considering an initial probability distribution of the prevalence. Our numerical studies illustrate the usefulness of the proposed index in ecological modeling. 

We note that our proposed index is defined on a larger range $(-\infty,1)$ while the other popular indices are mostly defined on a shorter interval. By appropriate scale adjustment we can also redefine it, but a larger range also allows a better comparison which is informative in bio-diversity. Also the choice of the initial probability distribution of the prevalence allows a bit flexibility in the proposed model.

In our current work, we did not come across any missing data. However, ecological data often contain a lot of missing information. Depending on the type of missingness observed in the data, one can either impute the missing values or can model the missingness. A simple 
data-augmentation technique might work for ignorable missingness. For non-ignorable missingness methods discussed in Daniels and Hogan (2008) can be implemented. We leave that as a different work.

\section*{Conflict of Interest}
The authors declare no conflict of interest

\section*{Funding Information}
There is no funding for the work presented in this paper.

\section*{References}

\noindent\hangindent=2em Blowes, S.A., Supp, S.R., Antão, L.H., et al. (2019), The geography of biodiversity change in marine and terrestrial assemblages, \emph{Science}, 366, 339--345.

\noindent\hangindent=2em Chiarucci, A., Fattorini, S., Foggi, B. et al. (2017), Plant recording across two centuries reveals dramatic changes in species diversity of a Mediterranean archipelago, \emph{Scientific Reports}, 7, 5415.

\noindent\hangindent=2em Crossley, N.A., Mechelli, A., Vértes, P.E. et al. (2013), Cognitive relevance of the community structure of the human brain functional coactivation network, \emph{Proceedings of the National Academy of Sciences}, 110, 11583--11588.

\noindent\hangindent=2em Daniels, M.J. and Hogan, J.W. (2008), \emph{Missing data in longitudinal studies: Strategies for Bayesian modeling and sensitivity analysis}, CRC press.

\noindent\hangindent=2em Demir, G.K. and Ozmehmet, K. (2005), Online local learning algorithms for linear discriminant analysis, 
\emph{Pattern Recognition Letters}, 26, 421--431.

\noindent\hangindent=2em Dornelas, M., Gotelli, N.J., McGill, B. et al. (2014), Assemblage time series reveal biodiversity change but not systematic loss, \emph{Science}, 344, 296--299.

\noindent\hangindent=2em Gianola, D., Manfredi, E. and Simianer, H. (2012), On measures of association among genetic variables,
\emph{Animal Genetics}, 43, 19--35.

\noindent\hangindent=2em Gotelli, N.J. and McCabe, D.J. (2002), Species co‐occurrence: a meta‐analysis of JM Diamond's assembly rules model, \emph{Ecology}, 83, 2091--2096.

\noindent\hangindent=2em Jaccard, P. (1912), The distribution of the flora in the alpine zone, \emph{New phytologist}, 11, 37--50.

\noindent\hangindent=2em Jolliffe, I.T. and Cadima, J. (2016), Principal component analysis: a review and recent developments, \emph{Philosophical transactions of the royal society A: Mathematical, Physical and Engineering Sciences}, 374, 20150202.

\noindent\hangindent=2em Kellman, M. and Schroder, T. (1983), The export similarity index: some structural tests, \emph{The Economic Journal}, 93, 193--198.

\noindent\hangindent=2em Kraft, N.J., Comita, L.S., Chase, J.M. et al. (2011), Disentangling the drivers of $\beta$ diversity along latitudinal and elevational gradients, \emph{Science}, 333, 1755--1758.

\noindent\hangindent=2em Mainali, K.P., Slud, E., Singer, M.C. and Fagan, W.F. (2022), A better index for analysis of co-occurrence and similarity, \emph{Science Advances}, 8, eabj9204.

\noindent\hangindent=2em Nazari, N., Mahmoodi, S. and Masihabadi, M.H. (2016), Employing diversity and similarity indices to evaluate Geopedological soil mapping in Miyaneh, East Azerbaijan Province, Iran, \emph{Open Journal of Geology}, 6, 1221.

\noindent\hangindent=2em Ocampo, P.S., Lázár, V., Papp, B., Arnoldini, M. (2014), Antagonism between bacteriostatic and bactericidal antibiotics is prevalent, \emph{Antimicrobial agents and chemotherapy}, 58, 4573--4582.

\noindent\hangindent=2em Oluyinka Christopher, A. (2020), Comparative analyses of diversity and similarity indices of west bank forest and block a forest of the International Institute of Tropical Agriculture (IITA) Ibadan, Oyo State, Nigeria, \emph{International journal of forestry research}, 1--8.

\noindent\hangindent=2em Plata, G., Henry, C.S. and Vitkup, D. (2015), Long-term phenotypic evolution of bacteria, \emph{Nature}, 517, 369--372.

\noindent\hangindent=2em Richer de Forges, B., Koslow, J.A. and Poore, G.C.B. (2000), Diversity and endemism of the benthic seamount fauna in the southwest Pacific, \emph{Nature}, 405, 944--947.

\noindent\hangindent=2em Simpson, E.H. (1949), Measurement of diversity, \emph{nature}, 163, 688--688.

\noindent\hangindent=2em Sorensen, T. (1948), A method of establishing groups of equal amplitude in plant sociology based on similarity of species content and its application to analyses of the vegetation on Danish commons, \emph{Biologiske skrifter}, 5, 1-34.

\noindent\hangindent=2em Tulloch, A.I., Chadès, I. and Lindenmayer, D.B. (2018), Species co-occurrence analysis predicts management outcomes for multiple threats, \emph{Nature ecology \& evolution}, 2, 465--474.

\newpage 
\section*{Appendix}

\begin{proof}
\textbf{(Proof of Theorem 1)}
 Let $\mu = \lambda_1(W^{-1}B)$ and $W = C^TC$ where $C$ is non-singular. Writing $C\Vec{x} = \Vec{y}$ we get 
 \begin{center}
     $\underset{\Vec{x} \not= \Vec{0}}{\text{max}} \frac{\Vec{x}^T B\Vec{x}}{\Vec{x}^T W \Vec{x}} = \underset{\Vec{y} \not= \Vec{0}}{\text{max}} \frac{\Vec{y}^T (C^{-1})^T B C^{-1} \Vec{y}}{\Vec{y}^T \Vec{y}}  = \lambda_1((C^{-1})^T B C^{-1})$
 \end{center}

Now the characteristic roots of $(C^{-1})^T B C^{-1}$ are all real and, by the next Theorem are the same as the characteristic roots of $ C^{-1}(C^{-1})^T B = W^{-1}B$  
Hence the theorem follows. Also, $\frac{\Vec{x}^T B \Vec{x}}{\Vec{x}^T W \Vec{x}} = \mu $ iff $C\Vec{x}$ is an eigenvector 
of $(C^{-1})^T B C^{-1}$ corresponding to $\mu$ which is equivalent to: $\Vec{x}$ is an 
eigenvector of $W^{-1}B$ corresponding to $\mu$. 
\end{proof}

\begin{theorem}
Let $A,B$ are $n \times n$ matrices and $A$ is non-singular (invertible). Then the characteristic polynomial of $AB$ and  $BA$ are the same.    
\end{theorem}

\begin{proof}
    Observe since $A$ is invertible, we can write 
    \begin{center}
        $A^{-1}(AB)A=BA$
 
    \end{center}
  So $AB$ and $BA$ are similar. Since similar matrices have the same characteristic polynomial( proved as the next theorem) we have characteristic polynomials of $AB$ and  $BA$ are the same. 
\end{proof}
\begin{theorem}
 Characteristic polynomials of similar matrices are the same.   
\end{theorem}

\begin{proof}
    Say, $A$ and $B$ are similar matrices. So, there exists nonsingular $P$ such that $A=PBP^{-1}$. So,

       $ det(A-\lambda I) = det(PBP^{-1}-\lambda I) = det(P(B-\lambda I)P^{-1})= det(P)det(B-\lambda I)det(P^{-1}) = det(B-\lambda I)$

\end{proof}

\end{document}